\documentclass[12pt]{article}
\usepackage{hyperref,slashed,graphicx,color,amsmath,amssymb}

\usepackage[numbers,sort&compress]{natbib}
\usepackage{subfig}

\begin{document}

\vskip 1cm
\begin{center}
{ \bf\large
Pulsar interpretation of lepton spectra measured by AMS-02
\vskip 0.2cm}
\vskip 0.3cm
{Jie Feng and Hong-Hao Zhang\footnote{Email: zhh98@mail.sysu.edu.cn}}
\vskip 0.3cm
{\it \small School of Physics and Engineering, Sun Yat-Sen University, Guangzhou
510275, China}
\end{center}
\vspace{0.3cm}
\begin{center}
{\bf Abstract}
\end{center}

AMS-02 recently published its lepton spectra measurement. The
results show that the positron fraction no longer increases above $\sim$200
GeV. The aim of this work is to investigate the possibility that the
excess of positron fraction is due to pulsars. Nearby known pulsars from ATNF catalogue are considered as a possible primary positron source of the high energy positrons. We find that the
pulsars with age $T\simeq (0.45\sim4.5)\times10^{5}$ yr
and distance $d<0.5$ kpc can explain the behavior of positron fraction
of AMS-02 in the range of high energy. We show that each of the four pulsars --- Geminga,
J1741-2054, Monogem and J0942-5552 --- is able to be a single source
satisfying all considered physical requirements. We also discuss the
possibility that these high energy $e{}^{\pm}$ are from multiple
pulsars. The multiple pulsars contribution predicts a positron fraction
with some structures at higher energies.

\newpage
\tableofcontents

\section{Introduction}

The positron fraction spectrum $e{}^{+}$/($e{}^{+}$+$e{}^{-}$)
in the cosmic ray (CR) contains two components: secondary $e{}^{\pm}$
produced by nuclei collision and primary $e^{-}$.
It is currently believed that these two components, each of which
will produce a diffused power low spectrum, predict a positron fraction
which goes down with energy. However, the latest results measured
by the Alpha Magnetic Spectrometer (AMS-02) with high accuracy indicate
that the positron fraction increases with energy above $\sim$8
GeV and does not increase with energy above $\sim$200 GeV \cite{Accardo:2014lma,Aguilar:2014mma}.
This {}``increasing'' behavior, which is also observed by the payload for antimatter
matter exploration and light-nuclei astrophysics (PAMELA) \cite{Adriani:2008zr,Adriani:2011cu,Adriani:2014xoa}
and the Fermi Large Area Telescope (Fermi-LAT) \cite{Grasso:2009ma,FermiLAT:2011ab}, is
not compatible with only diffused power low components. The {}``cutoff''
behavior above 200 GeV , which can be well described by a common source term with an exponential cutoff parameter in the Eq.(1) of \cite{Accardo:2014lma}, indicates that potential sources produce the
exceed of electron and positron pairs.

AMS-02 \cite{Accardo:2014lma,Aguilar:2014mma} is a state-of-the-art astroparticle detector installed on the
International Space Station (ISS). It carries a Transition Radiation
Detector (TRD) and a Electromagnetic Calorimeter (ECAL). These two
sub-detectors provide independent proton/lepton identification, which
will achieve a much larger proton rejection power of AMS-02 compared
with PAMELA which has only one Electromagnetic Calorimeter for proton/lepton
identification using the 3D shower shape and Energy-Momentum match (E/P). Compared with Fermi-LAT, AMS-02 has a large magnet
which can identify charge sign of the particle. Thus, the contamination
of electrons (also called ``charge confusion'' in \cite{Accardo:2014lma})
in the positron sample of AMS-02 is much smaller that that of Fermi-LAT.
For the reasons given above, there is much less proton or charge confusion
contamination in AMS-02 measurement than that in PAMELA or Fermi-LAT.
Here, we only interpret AMS-02 result due to the lack of knowledge
of the contamination control in PAMELA and Fermi-LAT measurements.

The AMS-02's recent measurements of positron fraction \cite{Accardo:2014lma},
$e{}^{+}$ flux, $e{}^{-}$ flux \cite{Aguilar:2014mma} and ($e{}^{-}$+$e{}^{+}$) flux \cite{Aguilar:2014fea}were published. The
$e{}^{-}$ flux contains three components: primary $e{}^{-}$, secondary
$e{}^{-}$ and $e{}^{-}$ from unknown sources. The $e{}^{+}$ flux
contains only two components: $e{}^{+}$ from secondary production
and primary $e{}^{+}$ from sources to be indentified. To avoid the unnecessary uncertainty
of primary $e{}^{-}$, the $e{}^{+}$ flux seems to be an ideal spectrum
to study extra sources. However, there is an acceptance uncertainty
from the detector itself in the $e{}^{+}$ and $e{}^{-}$ fluxes.
This uncertainty in $e{}^{+}$ flux is strongly correlated with that
in $e{}^{-}$ flux \cite{Aguilar:2014mma}, especially at high energies. The
positron fraction can avoid this systematic uncertainty \cite{Accardo:2014lma}.
For example, one can clearly see a drop at the last point (350 GeV
$\sim$ 500 GeV) in the positron fraction but cannot tell
a drop at the last point (370 GeV $\sim$ 500 GeV) in the
$e{}^{+}$ flux due to its larger error bars. Therefore, positron
fraction is used to study extra sources while $e{}^{-}$ flux is used
to estimate the primary $e{}^{-}$ which will affect the denominator
of $e{}^{+}$/($e{}^{+}$+$e{}^{-}$).

Recent studies have proposed some interpretations, such as dark matter annihilation
or decay \cite{Bergstrom:2008gr,Cirelli:2008jk,Barger:2008su,Cholis:2008hb,Cirelli:2008pk,
Yuan:2013eja,Yin:2008bs,Lin:2014vja,Boudaud:2014dta}, supernova remnants (SNRs) \cite{Blasi:2009hv,Blasi:2009bd,Mertsch:2009ph,Kachelriess:2012ag,
Cholis:2013lwa,DiMauro:2014iia},
secondary production in the interstellar medium (ISM) \cite{Blum:2013zsa}
and pulsars \cite{Boulares1989,Atoian:1995ux,Aharonian:1995zz,Chi:1995id,Hooper:2008kg,Yuksel:2008rf,
Profumo:2008ms,Malyshev:2009tw,Delahaye:2010ji,Pochon2010,Kashiyama:2010ui,Mertsch:2010fn,
Linden:2013mqa,Yin:2013vaa,Delahaye:2014osa,Lin:2014vja,Boudaud:2014dta}.
Cosmic ray flux data can also be together with other observations (like the dark matter relic density and the direct detection experimental results etc.) to give a combined constraint on dark matter models \cite{Zheng:2010js,Yu:2011by}.
Besides dark matter scenario,
the others can provide astrophysical explanations which do not require
the existence of new particles. SNRs model, for instance in \cite{Cholis:2013lwa} and \cite{Tomassetti:2015cva}, introduce some new mechanisms for the propagation model or special distributions of the primary sources. The {}``model-independent''
approach from \cite{Blum:2013zsa}, sets an upper limit of the positron
fraction by neglecting radiative losses of electrons and positron
but does not indicate any obvious cutoff in the spectrum. Among them,
the pulsar interpretation is one of the scenarios which predict a
cutoff at a few hundred GeV in the positron fraction spectrum and do not contridict other cosmic ray spectrums (eg. boron-to-carbon). The pioneering works on pulsar interpretaion of positron fraction have been performed by \cite{Hooper:2008kg,Profumo:2008ms,Yin:2013vaa} a few years ago. Combined analyses of the recent AMS-02 lepton data have been performed by \cite{DiMauro:2014iia} and \cite{Lin:2014vja}, with a global fit on positron fraction \cite{Accardo:2014lma}, $e{}^{+}$ flux, $e{}^{-}$ flux and ($e{}^{-}$+$e{}^{+}$) flux. To avoid the over-estimation of the $\chi{}^{2}$, however, only two out of four spectrums should be used in the fit. As the reasons given by the previous paragraph, we only study positron fraction and $e{}^{-}$ flux in this paper.

A pulsar is widely regarded as a rotating neutron star with a strong magnetosphere, which
can accelerate electrons, which will induce an electromagnetic cascade
through the emission of curvature radiation
\cite{Ruderman:1975ju,Shapiro:1983du,Cheng:1986qt,Contopoulos:2009vm}. This leads
to the production of high energy photons which eventually induces $e{}^{+}$$e{}^{-}$ pair production. This process produces the same amount of high energy $e{}^{+}$ and
$e{}^{-}$, which can escape from the magnetosphere and propagate
to the earth. There is a cutoff energy of the photons produced in
a pulsar, which leads to a cutoff in the positron fraction.

In this paper, DRAGON \cite{DragonWebsite,Evoli:2008dv,Gaggero:2013rya,Gaggero:2013nfa, Bernardo:2013036D} is used as a numerical tool
to model the propagation environment, to tune the related parameters
and to estimate the $e{}^{\pm}$ background. The authors
of \cite{Gaggero:2013rya,Gaggero:2013nfa, Bernardo:2013036D, Kissmann:2015kaa}
did a very complete work on three-dimensional cosmic-ray modeling.
In the 3-D models, they pointed out the spiral arms have
an effect on the propagation parameters. A 2-D model is
used in this paper because we focus on the lepton spectra
implication. Due to the energy loss of leptons, the effect
of spiral arms on the high energy leptons is less important than
that of the additional nearby sources contribution. ROOT is used to minimize
$\chi{}^{2}$ to get the best fit results. We consider six nearby
pulsars from ATNF catalogue \cite{ATNFwebsite,Manchester:2004bp} as the possible extra single
sources of the high energy positrons. We find only four, which are
Geminga, J1741-2054, Monogem and J0942-5552, can survive from all
considered physical requirements. We then discuss the possibility
that these high energy $e{}^{\pm}$ are from multiple pulsars. The
multiple pulsars contribution predicts a positron fraction with some
structures at higher energies.

The paper is organized as follows. Section \ref{sec-background} shows the way where $e{}^{\pm}$
background is estimated. In Section \ref{sec-single}, the properties of
pulsars are described and the profile of $e{}^{\pm}$ fluxes produced by
a pulsar is derived. The interpretation of positron fraction with one single pulsar is discussed in Section \ref{sec-single-2} and the hypothesis about multiple pulsars interpretion is tested in Section \ref{sec-multiple}. The conclusions are drawn
in Section \ref{sec-con}. In Appendix \ref{app-A}, the diffusion energy-loss equation for a burst-like source is solved with the spherically symmetric approximation.

\section{Propagation parameters and $e^{\pm}$ background \label{sec-background}}

The Galacitc background of the lepton fluxes are considered as three
main components, which are primary electrons from CR sources,
secondary electrons and positrons from the interactions between the
CR and the interstellar medium (ISM). The propagation of $e^{\pm}$ in the Galaxy obeys the following Ginzburg and Syrovatskii's equation \cite{Ginzburg2013origin}, also in \cite{Evoli:2008dv,Strong:2007nh}
\begin{eqnarray}
&&\frac{\partial f_i}{\partial t}-\nabla\cdot[(D\nabla-\vec{v}_c)f_i]
  -\frac{\partial}{\partial p}p^2D_{pp}\frac{\partial}{\partial p}\frac{f_i}{p^2}
  +\frac{\partial}{\partial p}\left[(\dot{p}-\frac{p}{3}\nabla\cdot\vec{v}_c)f_i\right]\nonumber\\
  &&=Q_i(\vec{x},t,p)+\sum_{j>i}c\beta n_{\mathrm{gas}}\sigma_{ji}f_j
  -c\beta n_{\mathrm{gas}}\sigma_{\mathrm{in}}f_i
\end{eqnarray}
where $p\equiv|\vec{p}|$ is the particle momentum; $f_i(\vec{x},t,p)$ is the particle number density of a species $i$ per unit momentum interval; $\vec{v}_c$ is the convection velocity, $\beta\equiv v/c$ is the ratio of velocity to the speed of light; $\sigma_{\mathrm{in}}$ is the total inelastic cross section onto the ISM gas, whose density is $n_{\mathrm{gas}}$; $\sigma_{ji}$ is the production cross section of the species $i$ by the fragmentation of the species $j$ (with $j>i$); and $Q_i(\vec{x},t,p)$ is the source term of species $i$, which can be thought to be steady $Q_i=Q_i(\vec{x},p)$ for background CR particles.

The spatial diffusion coefficient $D$ in the cylindrical coordinate system $(r,z)$ may be parameterized as \cite{Evoli:2008dv,Strong:2007nh,DiBernardo:2010is}
\begin{equation}
\left\{
\begin{array}{r@{\;=\;}l}
D(\rho,r,z) & D(\rho,r) e^{-|z|/z_t}\quad\text{or}\quad D(\rho,r,z)=D(\rho,r) (-L<z<L)\\
D(\rho,r) & D_0f(r)\beta\left(\frac{\rho}{\rho_0}\right)^\delta \label{spatial-diffusion-coe}
\end{array}
\right.
\end{equation}
where $\rho\equiv pc/(Ze)$ is defined as the particle magnetic rigidity, $z_t$ is the scale height of the diffusion coefficient, $L$ is the halo size, and $\delta$ is the index of the power-law dependence of the diffusion coefficient on the rigidity. $D_0$ is the normalization of the diffusion coefficient at the reference rigidity $\rho_0=4$ GV. Previous DRAGON papers \cite{Gaggero:2013rya,Gaggero:2013nfa, Bernardo:2013036D} tested a few models with the exponential profile, i.e., the left formula in \eqref{spatial-diffusion-coe}, which is more physical than the constant one, i.e., the right one. The effect of choosing different profiles on the electron and positron background is small if the parameters are properly set. In this paper, the constant profile is used in order to compute the pulsar profile in an analytical way, i.e., eq. \eqref{eq:e_flux} in Section \ref{sec-single}. The function $f(r)$ describes a possible radial dependence of $D$, and it can be taken to be unity for simplicity.

The diffusion coefficient in momentum space $D_{pp}$ is related to the spatial diffusion coefficient $D$ by \cite{Yuan:2013eja,DiBernardo:2010is,Zhang:2008tb,Seo1994}
\begin{eqnarray}
D_{pp}D=\frac{4p^2v_A^2}{3\delta(4-\delta^2)(4-\delta)w}
\end{eqnarray}
where $v_A$ is the Alfven velocity, and $\delta$ is the power-law index as given in \eqref{spatial-diffusion-coe}. $w$ is the ratio of magnetohydrodynamic wave energy density to the magnetic field energy density, and it is usually taken to be 1.

\begin{figure}[!htbp]
\includegraphics[width=1\textwidth]{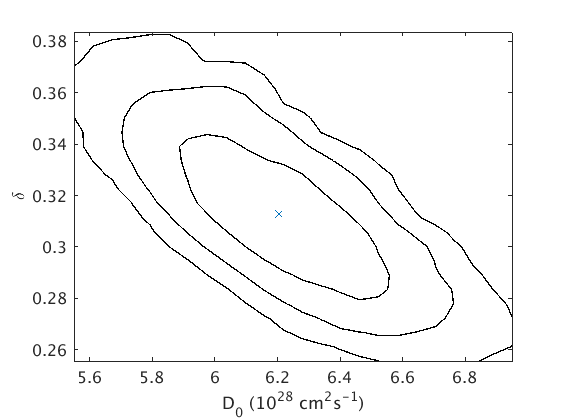}\caption{
Contour in the $D_{0}$ and $\delta$ plane. The cross shows the best fit value while the three closed curves from inside to outside show the 68.3\% C.L., 95.4\% C.L. and 99.7\% C.L. respectively.}
\label{Fig:D_delta}%
\end{figure}

DRAGON \cite{DragonWebsite,Evoli:2008dv,Gaggero:2013rya} is used to tune the propagation parameters according
to the B/C ratio, which is sensitive to the parameters. The Markov Chain Monte Carlo algorithm (MCMC, \cite{Lewis:2002ah}) is used to determine $D_{0}$ and $\delta$. The priors are shown in Table.~\ref{Tab:priors}. The posterior distributions can be shown in a contour in the $D_{0}$ and $\delta$ plane in Fig.~\ref{Fig:D_delta}.
\begin{table}[!htbp]
\centering
\caption{The priors of $D_{0}$ and $\delta$ }
\begin{tabular}{|c|c|c|c|}
\hline
 & start value & minimum value & maximum value \tabularnewline
\hline
\hline
$ D_{0}(\times10^{28}~ \mathrm{cm}^{2}\mathrm{s}^{-1}) $ & 3 & 2.5 & 7.5 \tabularnewline
\hline
$\delta$ & 0.40 & 0.20 & 0.65 \tabularnewline
\hline
\end{tabular}
\label{Tab:priors}%
\end{table}

The parameters with their 68\% C.L. uncertainties from the fit are as follows:

\begin{equation}
\left\{
\begin{array}{r@{\;=\;}l}
D_{0}\big|_{\rho_{0} = 4~\mathrm{GV}} & ( 6.20 \pm 0.31 ) \times10^{28}~ \mathrm{cm}^{2}\mathrm{s}^{-1}\\
		\delta & 0.31 \pm 0.03\\
		L & 4 ~\mathrm{kpc}\\
		v_{A} & 40~\mathrm{km}/\mathrm{s}
		\end{array}
		\right.
		\end{equation}
where the halo size $L$ is taken from the MED model of \cite{Donato:2003xg}, and the $v_{A}$ is fixed. These parameters are consistent with what the authors of Ref.\cite{Lin:2014vja} has got in the reaccelaration propagation model.
		
\begin{figure}[!htbp]
\subfloat[]{\includegraphics[width=0.5\textwidth]{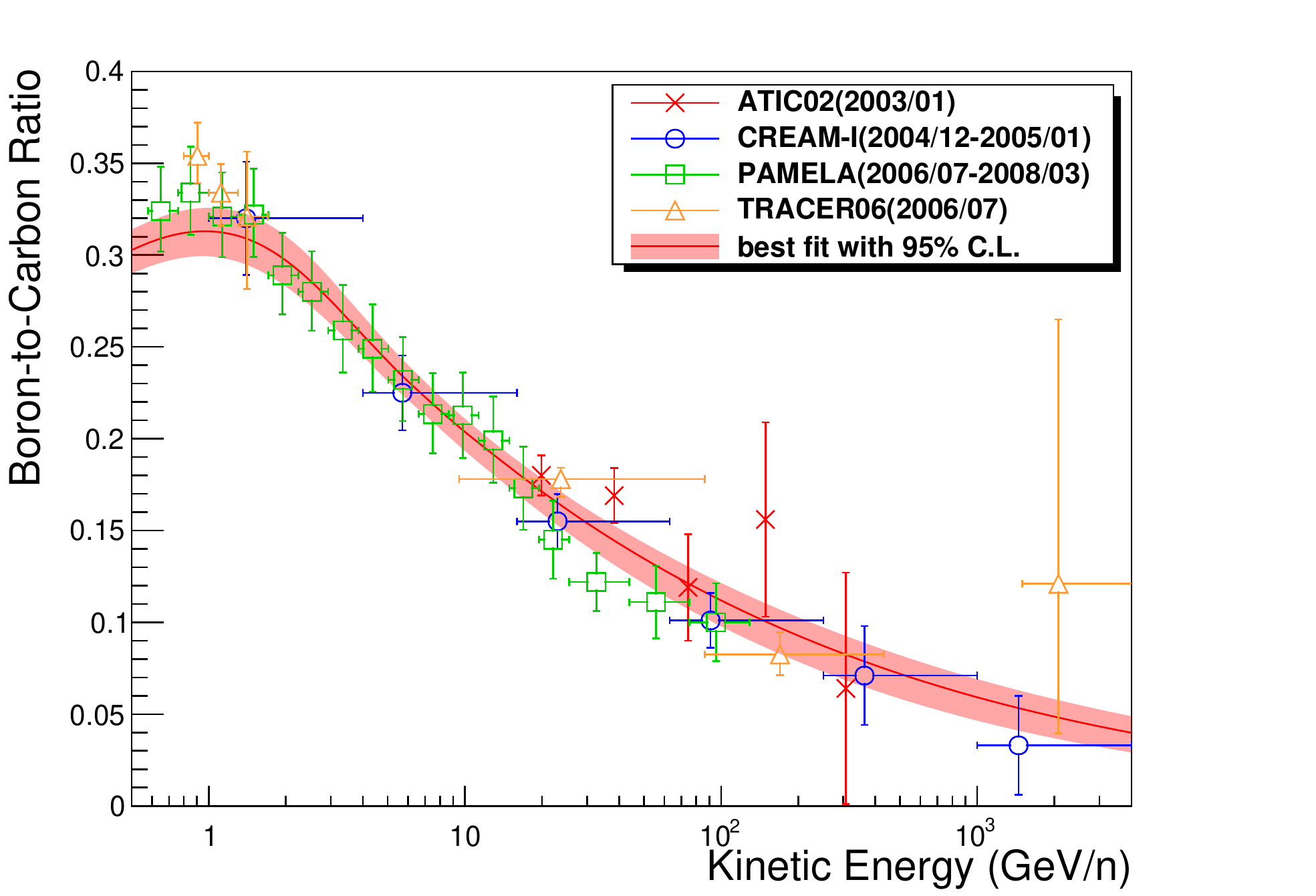}}
\subfloat[]{\includegraphics[width=0.5\textwidth]{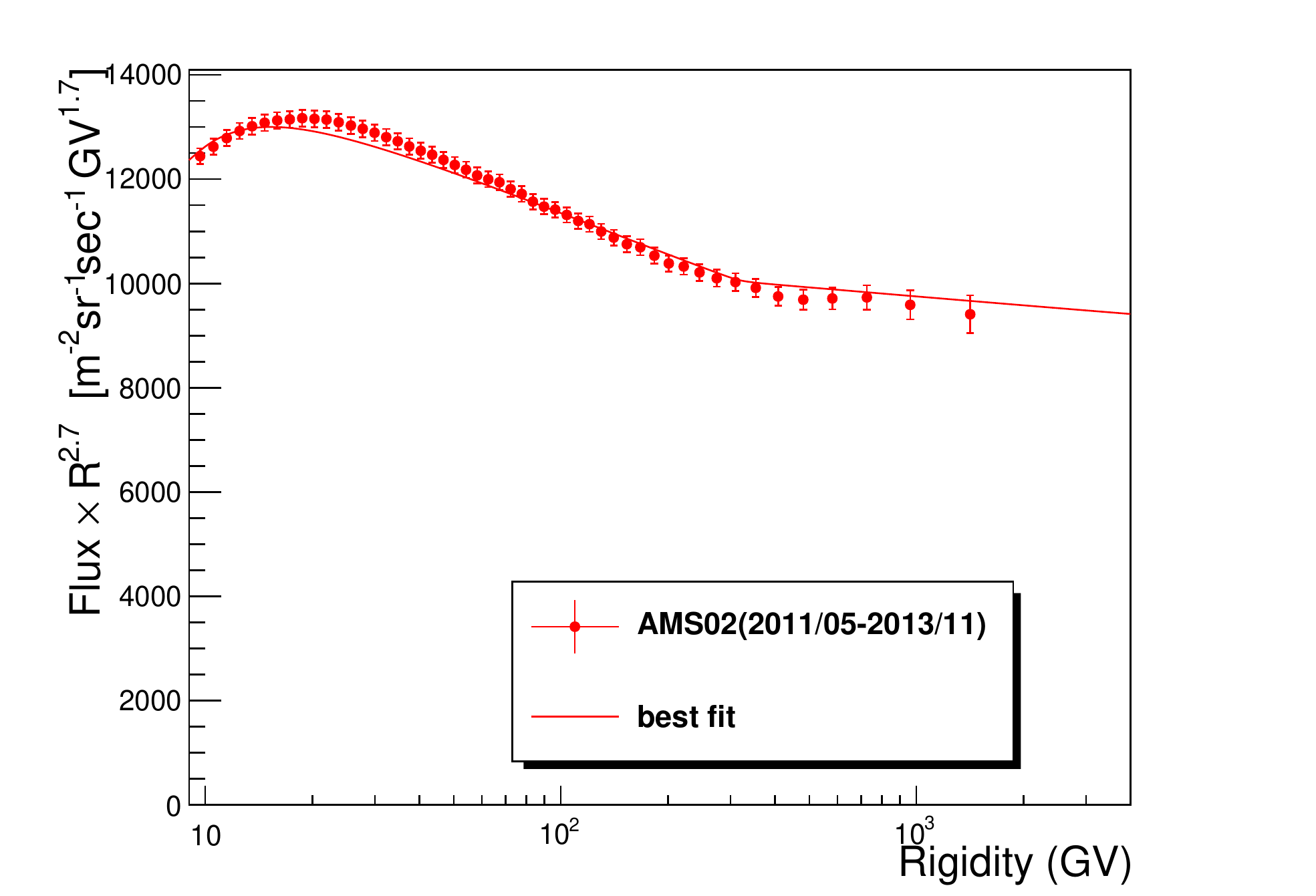}}
\caption{(a) model prediction of B/C ratio compared with measurements from
	PAMELA \cite{Adriani:2014xoa}, ATIC02 \cite{Panov:2007fe},
	       CREAM-I \cite{Ahn:2008my} and TRACER06 \cite{Obermeier:2012vg}.
		       (b) model prediction of proton flux compared with measurements from AMS-02 \cite{Aguilar:2015ooa}. The solar modulation
		       is taken as 500 MeV here. The red band in (a) shows
		       the variation of the propagation parameters $D_{0}$ and $\delta$ within 95\% C.L.
}
\label{Fig:BC_proton}%
\end{figure}

To avoid the uncertainty of solar modulation, AMS-02 proton flux \cite{Aguilar:2015ooa}
above 45 GV is fitted to get the injection spectra using MCMC \cite{Lewis:2002ah}.
Three breaks, which are 6.7 GV, 11 GV and 316($\pm$148) GV, are introduced
in the injection spectrum of nuclei. The proton spectral indice below and
above the breaks are 2.25, 2.35, 2.501($\pm$0.010) and 2.501-0.084($\pm$0.050), respectively.
The high energy spectral indices of helium, carbon and oxygen are shifted by -0.1 w.r.t those of proton
according to proton-to-helium ratio \cite{Adriani:2011cu}. The Ferriere
model \cite{Ferriere:2001rg} is used as the source distribution for the primary
components, e.g. SNRs for SNe type II.
To assure that the propagation parameters are correct, we need to compare
the model prediction with the boron-to-carbon ratio \cite{Adriani:2014xoa,Panov:2007fe, Ahn:2008my, Obermeier:2012vg} and the proton flux \cite{Aguilar:2015ooa}.
As shown in Fig.~\ref{Fig:BC_proton}, the set of
parameters used can reproduce the boron-to-carbon ratio and the proton flux well.
According to this set of parameters, we can obtain the fluxes of the secondary positrons and electrons.
A power-law spectrum with two breaks is introduced to parameterize
the injection spectrum of the primary electrons as a function of rigidity,
    \begin{equation}
    Q(\rho)\propto\begin{cases}
    \begin{array}{ll}
    (\rho/\rho_{\mathrm{br1}}^{e})^{-\gamma_{1}} &\quad (\rho<\rho_{\mathrm{br1}}^{e})\\
	    (\rho/\rho_{\mathrm{br1}}^{e})^{-\gamma_{2}} &\quad
	    (\rho_{\mathrm{br1}}^{e}\leq \rho\leq \rho_{\mathrm{br2}}^{e})\\
		    (\rho_{\mathrm{br2}}^{e}/\rho_{\mathrm{br1}}^{e})^{-\gamma_{2}}\cdot
		    (\rho/\rho_{\mathrm{br2}}^{e})^{-\gamma_{3}} &\quad
(\rho>\rho_{\mathrm{br2}}^{e})
	\end{array}
	\end{cases}\label{eq:1}
	\end{equation}
	The parameters are adjusted according to the electron flux from AMS-02 \cite{Aguilar:2014mma}.
	The agreement between the model and the data is shown in Section \ref{sec-multiple}.
	These spectral indices are $\gamma_{1}=1.95$, $\gamma_{2}=2.75$
	and $\gamma_{3}=2.5$ respectively. The breaks are $\rho_{\mathrm{br1}}^{e}=8.6$ GV
	and $\rho_{\mathrm{br2}}^{e}=110$ GV. Since the high energy breaks of primary particles,
	such as protons and helium, are found by PAMELA \cite{Adriani:2011cu} and recently
	confirmed by AMS-02 \cite{Aguilar:2015ooa}, it is reasonable to assume that there is
	also a high energy break in primary electron flux. More detailed discussion on the necessity of
	the high energy break $\rho_{\mathrm{br2}}^{e}$ can be found in \cite{Lin:2014vja}
	and \cite{Li:2014csu}, where the high energy break hypothesizes are in favor compared to the
	no-break ones. Ref.~\cite{Li:2014csu} gave us an estimation by taking the primary electron
	flux as $\Phi_{e-}-\Phi_{e+}$ and could roughly determine the break.

	\section{e$^{\pm}$ from a single pulsar \label{sec-single}}

	The pulsars are potential sources which could produce primary e$^{\pm}$
	at high energy \cite{Hooper:2008kg,Yuksel:2008rf,Profumo:2008ms,Malyshev:2009tw,Yin:2013vaa}. Electrons can be accelerated by the strong
	magnetosphere of the pulsars, and this acceleration produces photons. When those photons
	annihilate with each other, they can produce e$^{\pm}$ pairs. Thus, the e$^{\pm}$ energies
	are related to the pulsar magnetosphere. Assuming the pulsar magnetosphere
	as a magnetic dipole, this magnetic dipole radiation energy is proportional
	to the spin down luminosity. Due to this spin down (\textit{i.e.} slowing of rotation), the rotational
	frequency of a pulsar $\Omega\equiv 2\pi/P$ (with $P$ being the period) is
	a function of time as follows \cite{Hooper:2008kg,Profumo:2008ms,Yin:2013vaa}
	\begin{equation}
	\Omega(t)=\frac{\Omega_{0}}{\sqrt{1+t/\tau_{0}}},\label{eq:rotational_fre}
	\end{equation}
	where $\Omega_{0}$ is the initial spin frequency of the pulsar and
	$\tau_{0}$ is a time scale which describes the spin-down luminosity
	decays. $\tau_{0}$ cannot be directly obtained from pulsar timing observations, and it is assumed to be \cite{Profumo:2008ms,Yin:2013vaa}
	\begin{equation}
	\tau_{0}\simeq 10^{4}~\mathrm{yr}
	\end{equation}
	The rotational energy of the pulsar is $E(t)=(1/2)I\Omega^{2}(t)$.
	Here $I$ is the moment of inertia, which is related to the mass
	and the radius of the pulsar and can be regarded as a time independent
	value. The magnetic dipole radiation energy is equal to the energy
	loss rate,
	\begin{equation}
	|\dot{E}(t)|=I\Omega(t)\dot{|\Omega}(t)|
	=\frac{I\Omega_{0}^{2}}{2}\frac{1}{\tau_{0}(1+t/\tau_{0})^{2}}\label{eq:Edot}
	\end{equation}
	The total energy loss of a pulsar is \cite{Profumo:2008ms,Malyshev:2009tw,Yin:2013vaa}
	\begin{equation}
	E_{\mathrm{tot}}(t)=\int_0^t dt'|\dot{E}(t')|
	=\frac{I\Omega_{0}^{2}}{2}\frac{t/\tau_{0}}{1+t/\tau_{0}}
	=|\dot{E}(t)|t\left(1+\frac{t}{\tau_{0}}\right)\label{eq:E_total}
	\end{equation}
	The total energy injection of $e^{\pm}$ out of a pulsar should be
	proportional to the total energy loss
	\begin{equation}
	E_{\mathrm{out}}(t)=\eta E_{\mathrm{tot}}(t)=\eta|\dot{E}(t)|t\left(1+\frac{t}{\tau_{0}}\right)\label{eq:E_out}
	\end{equation}
	where $\eta$ is the efficiency of the injected $e^{\pm}$ energy
	converted from the magnetic dipole radiation energy.

	The pulsar characteristic age is defined as \cite{Manchester:2004bp}
	\begin{equation}
	T\equiv\frac{P}{2\dot{P}}=\frac{\Omega}{2|\dot{\Omega}|}=t+\tau_0
	\end{equation}
	For a mature pulsar with $t\gg \tau_0$, we have $T\simeq t$. In this condition, eqs.~\eqref{eq:Edot}, \eqref{eq:E_total} and \eqref{eq:E_out} become
	\begin{eqnarray}
	&&|\dot{E}(T)|\simeq \frac{I\Omega_{0}^{2}}{2}\frac{\tau_0}{T^2}\\
		&&E_{\mathrm{tot}}(T)\simeq |\dot{E}(T)|\frac{T^2}{\tau_0}\label{E-tot-2}\\
		&&E_{\mathrm{out}}(T)\simeq \eta |\dot{E}(T)|\frac{T^2}{\tau_0}\label{E-out-2}
		\end{eqnarray}

		The propagation equation for
		the $e^{\pm}$ can be described as \cite{Hooper:2008kg,Yin:2013vaa}
		\begin{equation}
		\frac{\partial f}{\partial t}=D(E)\nabla^2f+\frac{\partial}{\partial E}[b(E)f]+Q(\vec{x},t,E),\label{eq:propagation_eq}
		\end{equation}
		where $f(\vec{x},t,E)$ is the number density per unit energy interval of $e^{\pm}$; $D(E)=(v/c)D_{0}(E/4~\mathrm{GeV})^{\delta}$
		is the diffusion coefficient with the velosity $v$ of the particle,
		the speed $c$ of light, $D_{0}$ and $\delta$ the same as the parameters
		used to calculate the background in Section~\ref{sec-background}; and $b(E)\equiv-dE/dt=b_0E^{2}$
		with $b_{0}=1.4\times10^{-16}~\mathrm{GeV}^{-1}\mathrm{s}^{-1}$ is the rate of energy
		loss due to inverse Compton scattering and synchrotron \cite{Grasso:2009ma,Profumo:2008ms,Yin:2013vaa}.

		The source term $Q(\vec{x},t,E)$ of a pulsar can be described by
		a burst-like source with a power-law energy spectrum and an exponential
		cutoff
		\begin{equation}
Q(\vec{x},t,E)=Q_0E^{-\alpha}\exp\left(-\frac{E}{E_{\mathrm{cut}}}\right)
	\delta^3(\vec{x}-\vec{x}_0)\delta(t-t_0),
	\label{eq:Q_source}
	\end{equation}
	where $Q_{0}$ is the normalization factor related to the total injected energy $E_{\mathrm{out}}$,
	$\alpha$ is the spectral index, and $E_{\mathrm{cut}}$ is the cutoff energy.

	In Appendix~\ref{app-A},
 we briefly review how to solve the equation \eqref{eq:propagation_eq} with the source \eqref{eq:Q_source}. The method is equivalent to many previous works (for example, Refs.~\cite{Atoian:1995ux,Yin:2013vaa}).
 Using the results, \textit{i.e.}
	eqs.~\eqref{solution-final} and \eqref{diff-distance-final}, in Appendix~\ref{app-A}, we obtain the electron or positron flux observed at the earth as follows:
	\begin{equation}
	\Phi_{e}(r,t_{\mathrm{dif}},E)=\frac{c}{4\pi}f
	=\frac{c}{4\pi}\frac{Q_0E^{-\alpha}}{\pi^{\frac{3}{2}}r_{\mathrm{dif}}^3}
	\left(1-\frac{E}{E_{\mathrm{max}}}\right)^{\alpha-2}
	\exp\left[-\frac{E/E_{\mathrm{cut}}}{(1-E/E_{\mathrm{max}})}
	-\frac{d^2}{r_{\mathrm{dif}}^2}\right],\label{eq:e_flux}
	\end{equation}
	where $d$ is the distance between the earth and
	the source, the diffusion distance $r_{\mathrm{dif}}$ is given by
	\begin{equation}
	r_{\mathrm{dif}}(t_{\mathrm{dif}},E)=
	2\sqrt{\frac{D(E)t_{\mathrm{dif}}}{(1-\delta)}
		\frac{E_{\mathrm{max}}}{E}
		\left[1-\left(1-\frac{E}{E_{\mathrm{max}}}\right)^{1-\delta}\right]}\label{eq:r_dif}
		\end{equation}
		and the diffusion time $t_{\mathrm{dif}}$ is  the time a charged particle travels
		in the ISM before it reaches the earth. The electrons and positrons may be trapped in
		the pulsar wind nebula (PWN) for some time before they escape. The age of a pulsar is
		$T = t_{\mathrm{escape}} + t_{\mathrm{dif}}$, where $t_{\mathrm{escape}}$ is the time
		before the leptons escape from the PWN. In some case, $t_{\mathrm{escape}}$ and
		$t_{\mathrm{dif}}$ can be of the same order of magnitude, and then the discussion will
		be complicated. In some other case, $t_{\mathrm{escape}}$ could be negligible. For instance,
		when the SNR is evolving into the "Sedov-Taylor" phase, the leptons in it are trapped
		 (See Ref.~\cite{Gaensler:2006ua} and references there in). In that case,
		the time $t_{\mathrm{escape}}$, during which the SNR reverse shock collides with the PWN
		forward shock, is typically a few $10^{3}$ yr~\cite{Gaensler:2006ua}, which is small comparing to the
		ages of the pulsars we studied here, which are around $10^{5}$ yr. In this work, we consider
		the latter case and neglect $t_{\mathrm{escape}}$ for simplicity. We leave the case of
		large $t_{\mathrm{escape}}$ to a further specific study. Thus, we
		assume that $t_{\mathrm{dif}}\simeq T$. The maxium energy $E_{\mathrm{max}}$ is defined as
		\begin{equation}
		E_{\mathrm{max}}=1/(b_{0}T).\label{eq:E_max}
		\end{equation}

\begin{figure}[!htbp]
		\includegraphics[width=1\textwidth]{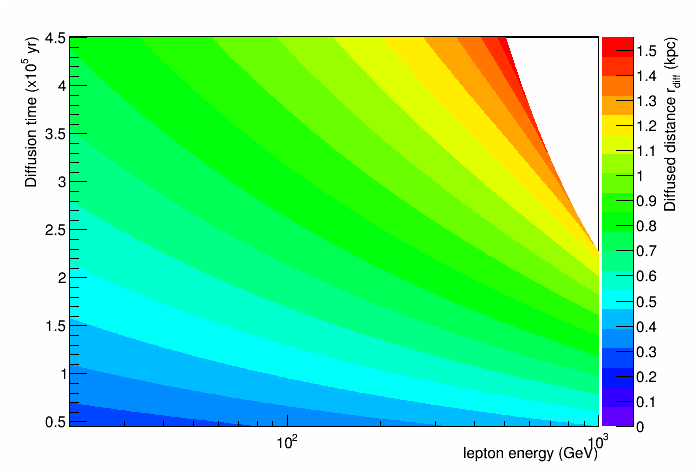}\caption{$r_{\mathrm{dif}}$ as a function of $t_{\mathrm{dif}}$ and $E$, which is from eq.~\eqref{eq:r_dif}. The lepton energy $E$
			is the $e{}^{+}$ (or $e^{-}$) energy detected at location away from the
				pulsar with the diffusion distance $r_{\mathrm{dif}}$. $r_{\mathrm{dif}}$ increases
				with $E$.}
				\label{Fig:r_d}%
				\end{figure}

		The positron fraction from AMS-02 implies a primary positron source with a cut-off
		energy $1/E_{s} = 1.84 \pm 0.58\:TeV^{-1}$ in their "minimal" model \cite{Accardo:2014lma},
		which corresponds to $E_{s} \in \left[490,790\right]$. Due to the limitation of statistics
		of high energy e- and e+ measured by AMS-02, the upper bound 790 GeV is not a strict limit.
		Thus, we consider a primary e+ and e- source contribution with a cut-off energy
		$E_{\mathrm{cut-off}}\simeq$ (500 $\sim$ 5000) GeV, which corresponds
		to a pulsar with an age $T\simeq (0.45\sim 4.5)\times10^{5}$ yr according to \eqref{eq:E_max}.
		The term $\exp\left[-\frac{d^2}{r_{\mathrm{dif}}^2}\right]$ in \eqref{eq:e_flux} tells us that
		a pulsar with $d>r_{\mathrm{dif}}$ requires a larger
		normalization $Q_0$, which hints a larger $E_{\mathrm{out}}$, a larger $\eta$ in \eqref{eq:E_out}, or both.
		$r_{\mathrm{dif}}>d$ is required in our study, whose physical interpretation is that the distance
		a particle travels in the ISM should be larger
		than the distance between the earth and the source.
		Eq.~\eqref{eq:r_dif} tells us that $r_{\mathrm{dif}}$ is as a function of diffusion time $t_{\mathrm{dif}}$
		and lepton energy $E$, as is shown by Fig.~\ref{Fig:r_d} where the color scale indicades $r_{\mathrm{dif}}$.
		For $T\simeq (0.45\sim 4.5)\times10^{5}$ yr and the lepton energy $E=1000 $ GeV,
		$r_{\mathrm{dif}}$ is always greater than 0.5 kpc.
		Selecting pulsars with $d<0.5 $ kpc and $T\simeq (0.45\sim 4.5)\times10^{5}$ yr, high engery leptons
		they produced can reach the earth.

				Thus, the pulsars with ages $T\simeq (0.45\sim 4.5)\times10^{5}$ yr
				and distance $d<0.5$ kpc can explain the behavior of positron fraction
				of AMS-02 at high energy range.

				\section{Single pulsar interpretation \label{sec-single-2}}

				A few simple examples using a single pulsar are given to explain high energy positron fraction
				of AMS-02 \cite{Accardo:2014lma}. The background electrons
				and positrons are described in Section~\ref{sec-background}. The primary electron flux
				is scaled by a normalization factor $A{}_{prim,e^{-}}$ since it is not possible to constrain the electron flux contribution from SNRs. The
				age $T$ and the distance $d$ are taken from the ATNF catalogue and the
				positron fraction is fitted to obtain the free parameters in \eqref{eq:e_flux}, the spectral index $\alpha$, and the normalization $Q_0$. $Q_0$ is fixed by the relation \cite{DiMauro:2014iia,Malyshev:2009tw,Linden:2013mqa} $E_{\mathrm{out}}=\int_{E_{\mathrm{min}}}^{E_{\mathrm{max}}} dEEQ(E)\simeq\int_0^\infty dEEQ(E)$, which approximately yields $Q_{0}\simeq E_{\mathrm{out}}$ for $\alpha\simeq 2$. The cutoff energy $E_{\mathrm{cut}}$ is set to be 5000~GeV, which is large enough, as it does not change the shape
				of pulsar contribution. Since we are interested in the positron excess at high energies, the fit is started from 10 GeV where the effect of solar modulation is negligible.

\begin{figure}[!htbp]
						    \subfloat[]{\includegraphics[width=0.5\textwidth]{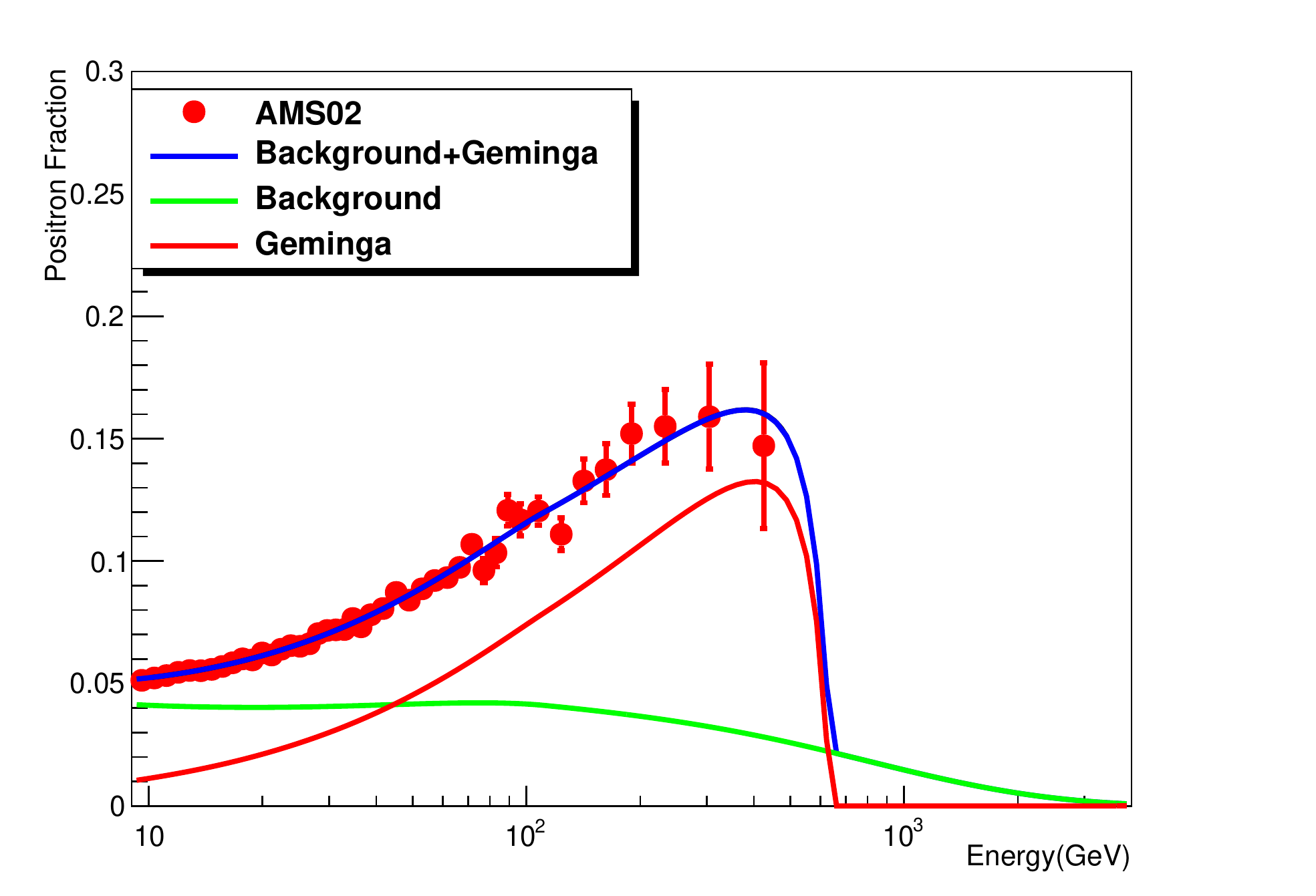}}
						    \subfloat[]{\includegraphics[width=0.5\textwidth]{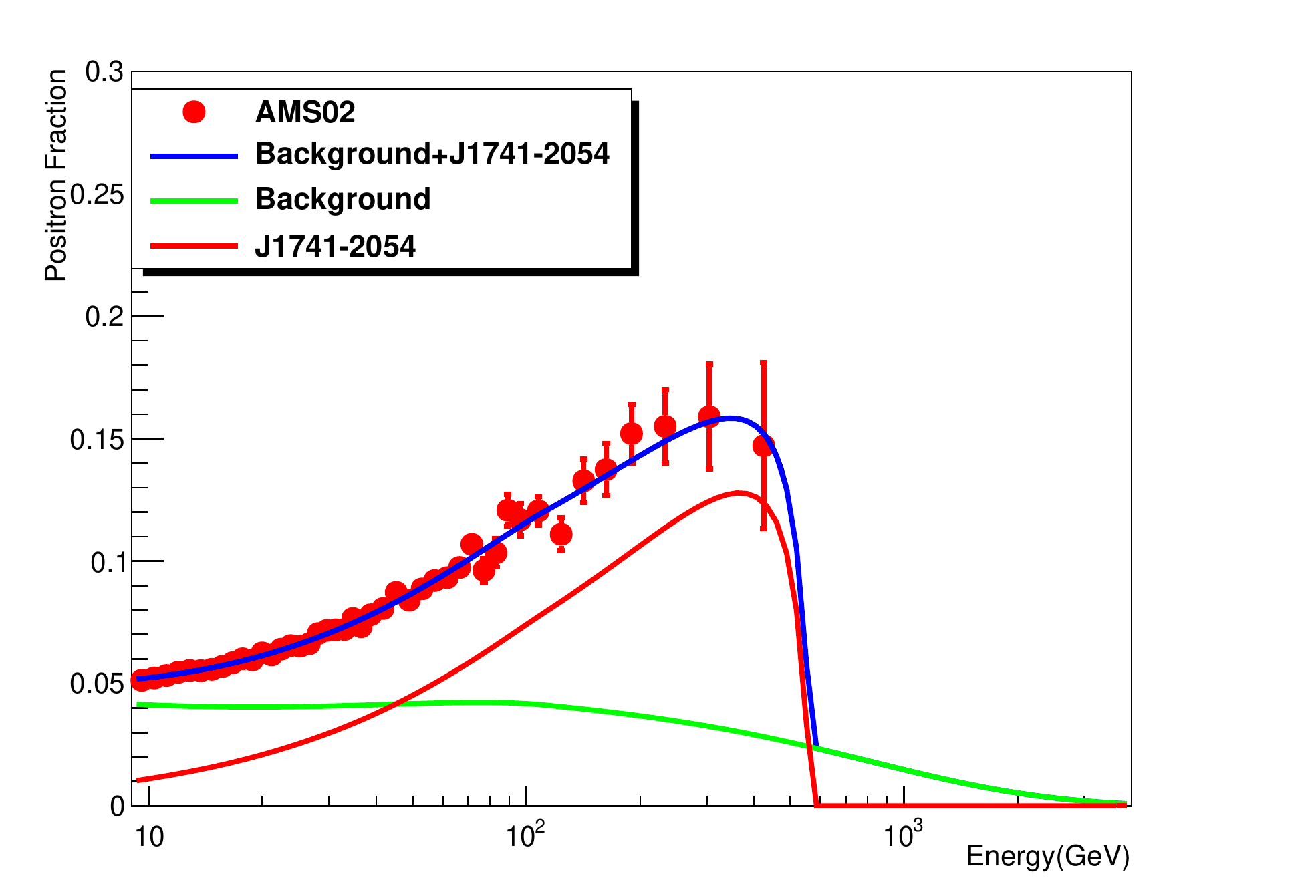}}

						    \subfloat[]{\includegraphics[width=0.5\textwidth]{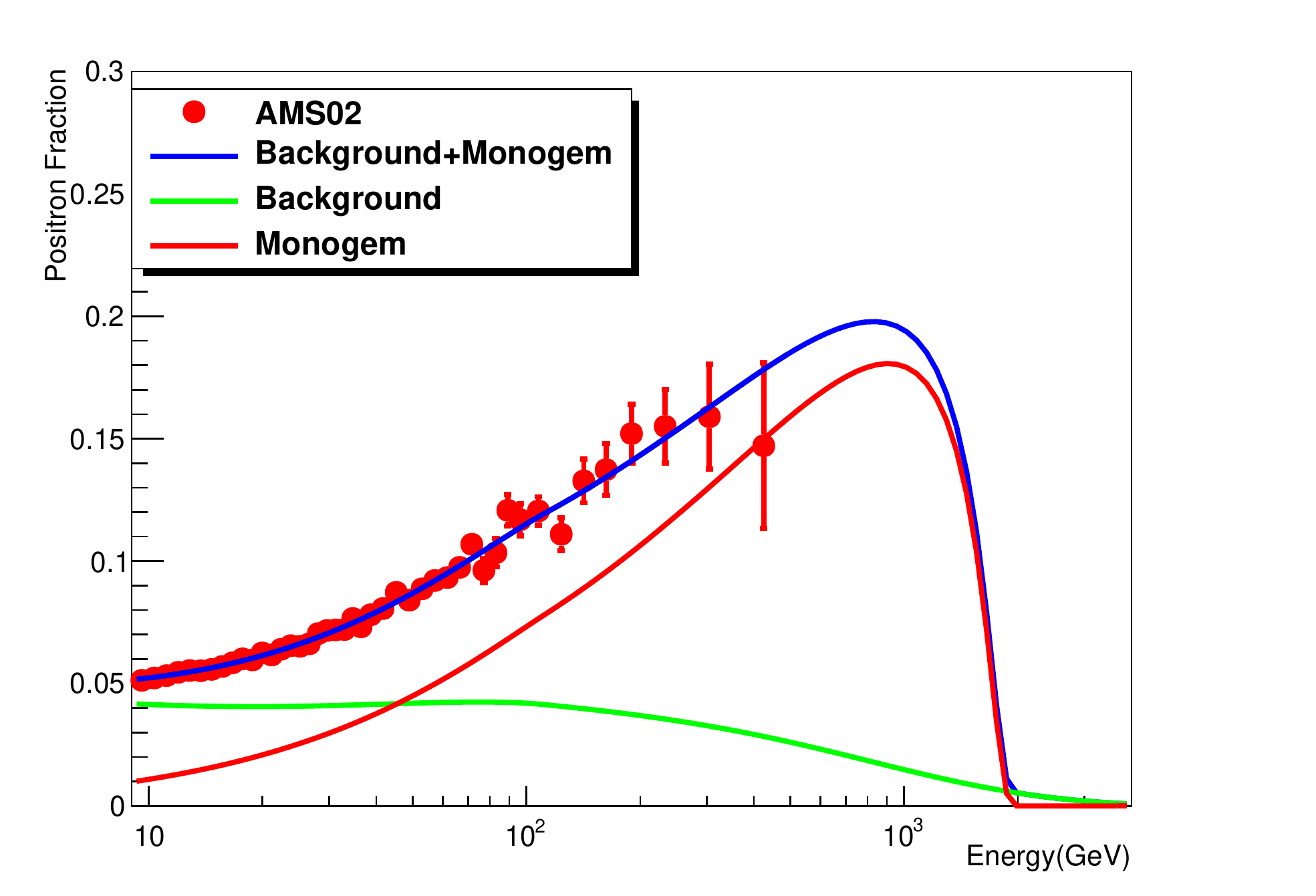}}
						    \subfloat[]{\includegraphics[width=0.5\textwidth]{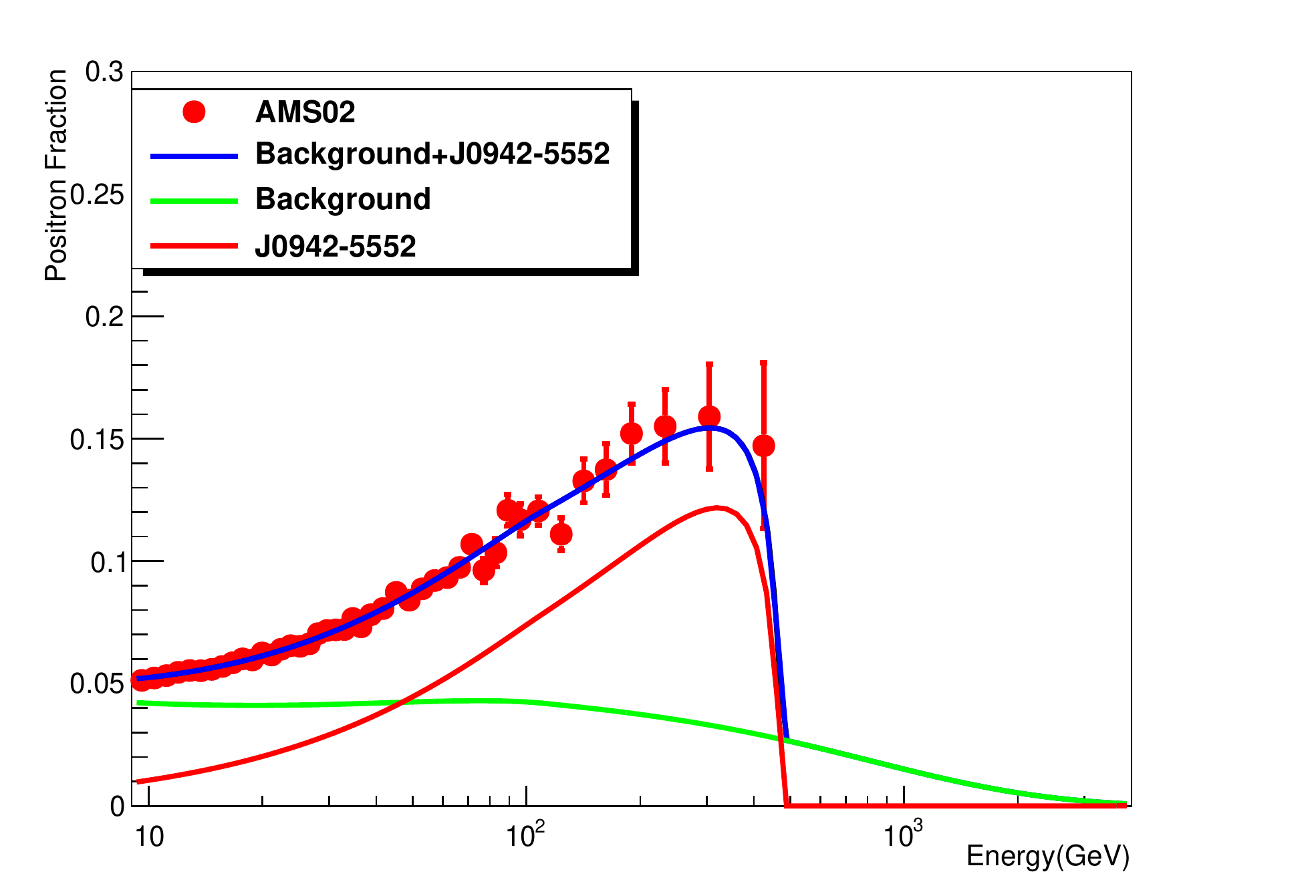}}

						    \subfloat[]{\includegraphics[width=0.5\textwidth]{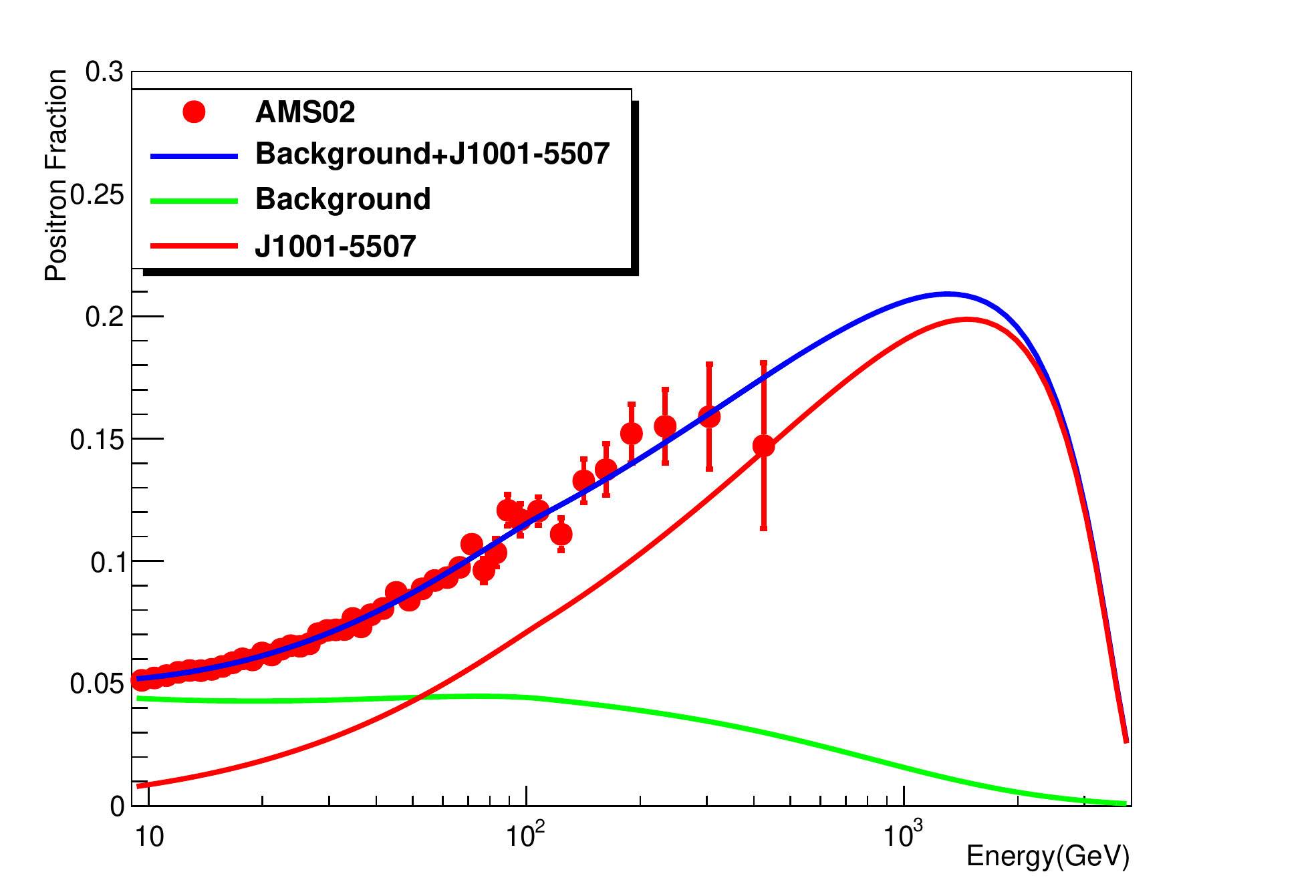}}
						    \subfloat[]{\includegraphics[width=0.5\textwidth]{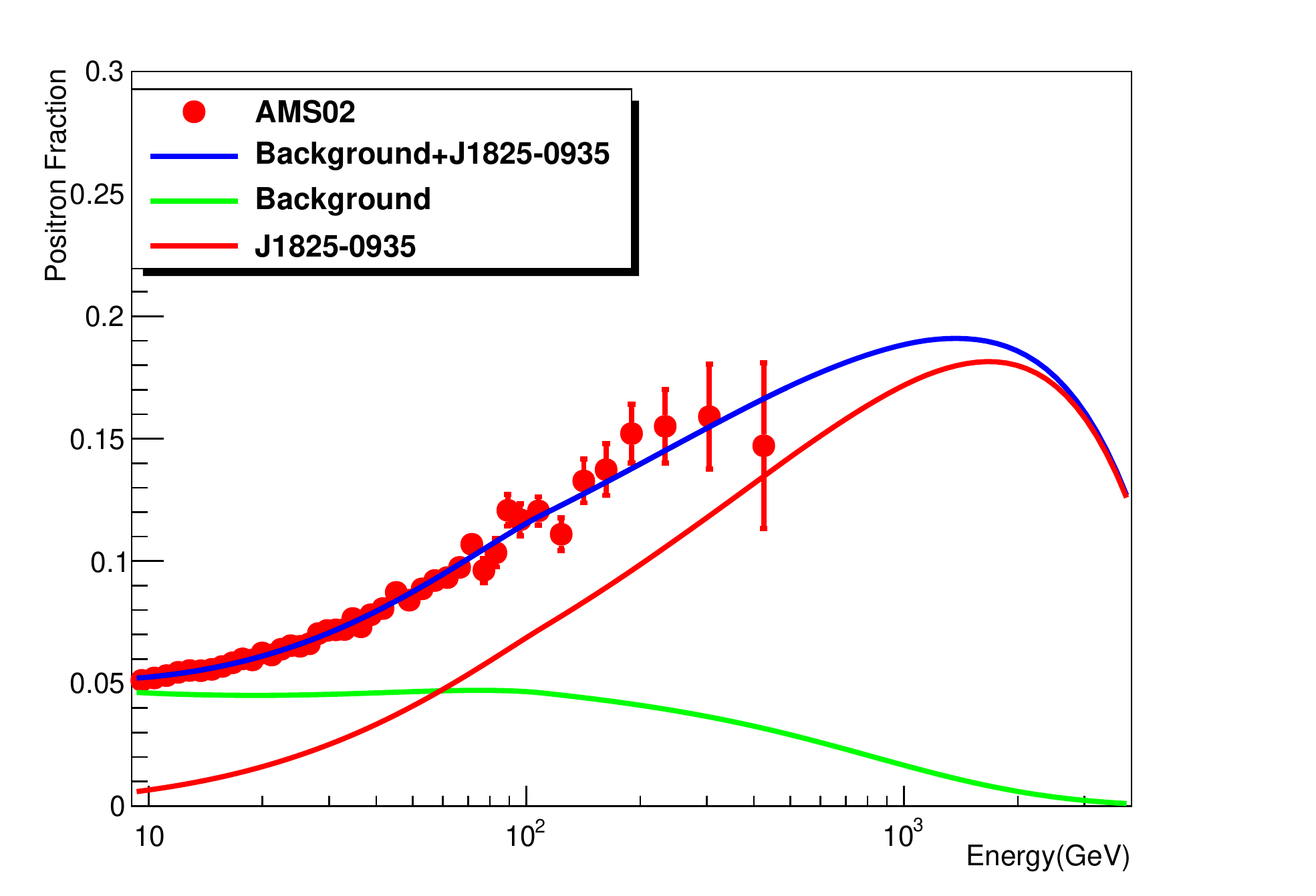}}
						    \caption{Single pulsar model can explain the positron fraction very well. According
							    to the fitting result, the spectral indices are almost the same. }
							    \label{fig3}
							    \end{figure}

				Six nearby single pulsars, whose ages $T\simeq (0.45\sim 4.5)\times10^{5}$ yr and distance $d<0.5$ kpc, are used to fit the positron fraction. Minuit
				package in ROOT is used to determine the parameters to minimize $\chi^{2}$. The best results of the single pulsars are listed in Table \ref{Tab:fit_parameters_of_single_pulsars}.
				\begin{table}[!htbp]
				\centering
				\begin{tabular}{|c|c|c|c|c|c|c|}
				\hline
				Pulsar name & $d$(kpc) & $T$($10{}^{5}$~yr) & $log{}_{10}(\frac{Q_{0}}{\mathrm{GeV}})$ & $\alpha$ & $A_{prim,e-}$ & $\chi^{2}/ndf$\tabularnewline
				\hline
				\hline
				Geminga & 0.25 & 3.42 & 50.5 & 2.04 & 0.50 & 26.8/40\tabularnewline
				\hline
				J1741-2054 & 0.25 & 3.86 & 50.6 & 2.03 & 0.50 & 26.8/40\tabularnewline
				\hline
				Monogem & 0.28 & 1.11 & 50.1 & 2.15 & 0.50 & 27.3/40\tabularnewline
				\hline
				J0942-5552 & 0.30 & 4.61 & 50.6 & 2.01 & 0.49 & 27.7/40\tabularnewline
				\hline
				J1001-5507 & 0.30 & 4.43 & 50.1 & 2.34 & 0.47 & 27.6/40\tabularnewline
				\hline
				J1825-0935 & 0.30 & 2.32 & 50.5 & 2.61 & 0.44 & 28.8/40\tabularnewline
				\hline
				\end{tabular}
				\caption{Parameters of six nearby single pulsars from the best fit results.
					The $\chi^{2}/ndf$ from the fits of Geminga, J1741-2054, Monogem,
					    J0942-5552 and J1001-5507 are smaller than 1, which show a good agreement
						    between those single pulsar models and the experiment. }
						    \label{Tab:fit_parameters_of_single_pulsars}%
						    \end{table}
						    The results are also shown in Fig.~\ref{fig3}.
							    Using the parameters of the best fit results, the positron fraction
							    can be well reproduced by these single pulsar's contributions. Table
							    \ref{Tab:fit_parameters_of_single_pulsars} tells us that the normalization
							    $A{}_{prim,e^{-}}$ are around 0.5 and the spectral
							    indices $\alpha$ of different pulsars are around 2.

							    We can estimate the injection efficiency $\eta$ from the pulsar.
							    Take Geminga as an example, the spin-down energy loss rate of Geminga
							    $|\dot{E}(T)|=3.2\times10^{34}~\mathrm{erg}/\mathrm{s}$. The total radiation energy of the
							    magnetic dipole can be derived from eq.~\eqref{E-tot-2} as $E_{\mathrm{tot}}(T)\simeq |\dot{E}(T)|T^2/\tau_0=1.2\times10^{49}~\mathrm{erg}$.
							    From the fit, we get the injection energy $E_{\mathrm{out}}/2=10{}^{50.5}~\mathrm{GeV}\simeq 5.19\times10^{47}$ erg.
							    From \eqref{E-out-2}, we get $\eta\sim8.7\%$. This efficiency is
							    consistent with the previous studies by \cite{Hooper:2008kg} and \cite{Yin:2013vaa}.
							    We can perform similar studies on the other five pulsars, whose results
							    are listed in Table \ref{Tab:injection_efficiency}.
							    \begin{table}[!htbp]
							    \centering
							    \begin{tabular}{|c|c|c|c|c|}
							    \hline
							    Pulsar name & $|\dot{E}|(10^{33}~\mathrm{erg})$ & $E{}_{\mathrm{tot}}(10^{49}~\mathrm{erg})$ & $E{}_{\mathrm{out}}(10^{47}~\mathrm{erg})$ & $\eta(\%)$\tabularnewline
							    \hline
							    \hline
							    Geminga & 32 & 1.2 & 5.19 & 8.7\tabularnewline
							    \hline
							    J1741-2054 & 9.5 & 0.46 & 5.83 & 25\tabularnewline
							    \hline
							    Monogem & 38 & 0.16 & 1.90 & 24\tabularnewline
							    \hline
							    J0942-5552 & 3.1 & 0.21 & 7.02 & 67\tabularnewline
							    \hline
							    J1001-5507 & 0.68 & 0.043 & 1.89 & 88\tabularnewline
							    \hline
							    J1825-0935 & 4.6 & 0.082 & 5.09 & 120\tabularnewline
							    \hline
							    \end{tabular}\caption{Electron injection efficiency $\eta$ of the six nearby pulsars. For
								    single pulsar interpretation of positron fraction, the results of Geminga,
									   J1741-2054, Monogem and J0942-5552 are thought to be reasonable while
										   the posibilities of J1001-5507 and J1825-0935 as the high energy positron
										   sources can be excluded. }
										   \label{Tab:injection_efficiency}%
										   \end{table}
										   A smaller $\eta$
										   means it is easier for this pulsar to produce the same amount of positrons
										   and electrons. The efficiency required by J1001-5507 or J1825-0935
										   is too large to satisfy the physics condition for single pulsar interpretation.
										   Geminga, J1741-2054, Monogem and J0942-5552 are the only candidates
										   which survive from our selection so far.\footnote{Considering that
the uncertainty of $log{}_{10}(\frac{Q_{0}}{\mathrm{GeV}})$ from the fit is $\pm0.1$,
the $E{}_{\mathrm{out}}$ for J1001-5507 is $1.89^{+0.44}_{-0.36}\times 10^{47}erg$ .
Thus, $\eta = 88^{+21}_{-17} $ \% for J1001-5507. There is no enough strong evidence that this $\eta$ is smaller than 1.
One should also note that $\eta = 67^{+14}_{-12} $ \% for J0942-5552, which is 2$\sigma$ smaller than 1.}

										   \section{Multiple pulsars interpretation \label{sec-multiple}}

										   The extra high energy positrons may come from serveral pulsars. We
										   perform similar study for multiple pulsars as we do for a single pulsar.
										   Benefiting from the
										   study in Section \ref{sec-single-2}, we can assume that the spectral indices $\alpha$
										   of all the pulsars are the same. Considering the physical models of the pulsars are similar,
										   we make another assumption that the electron injection efficiencies $\eta$ are the same.
										   These two assumptions help us reduce the number of
										   free parameters. The discussion on $\eta$ from single pulsar in
										   Section \ref{sec-single-2} tells us that Geminga, J1741-2054 and Monogem will give
										   a much larger contribution to the high energy positron than J0942-5552.
										   In other words, the $\eta$ of J0942-5552 in Table \ref{Tab:injection_efficiency}
										   is much larger than that of Geminga, which implies that the contribution from J0942-5552
										   in the multiple pulsars interpretation can be negligible compared
										   with that from Geminga.

\begin{figure}[!htbp]
										   \subfloat[]{\includegraphics[width=0.5\textwidth]{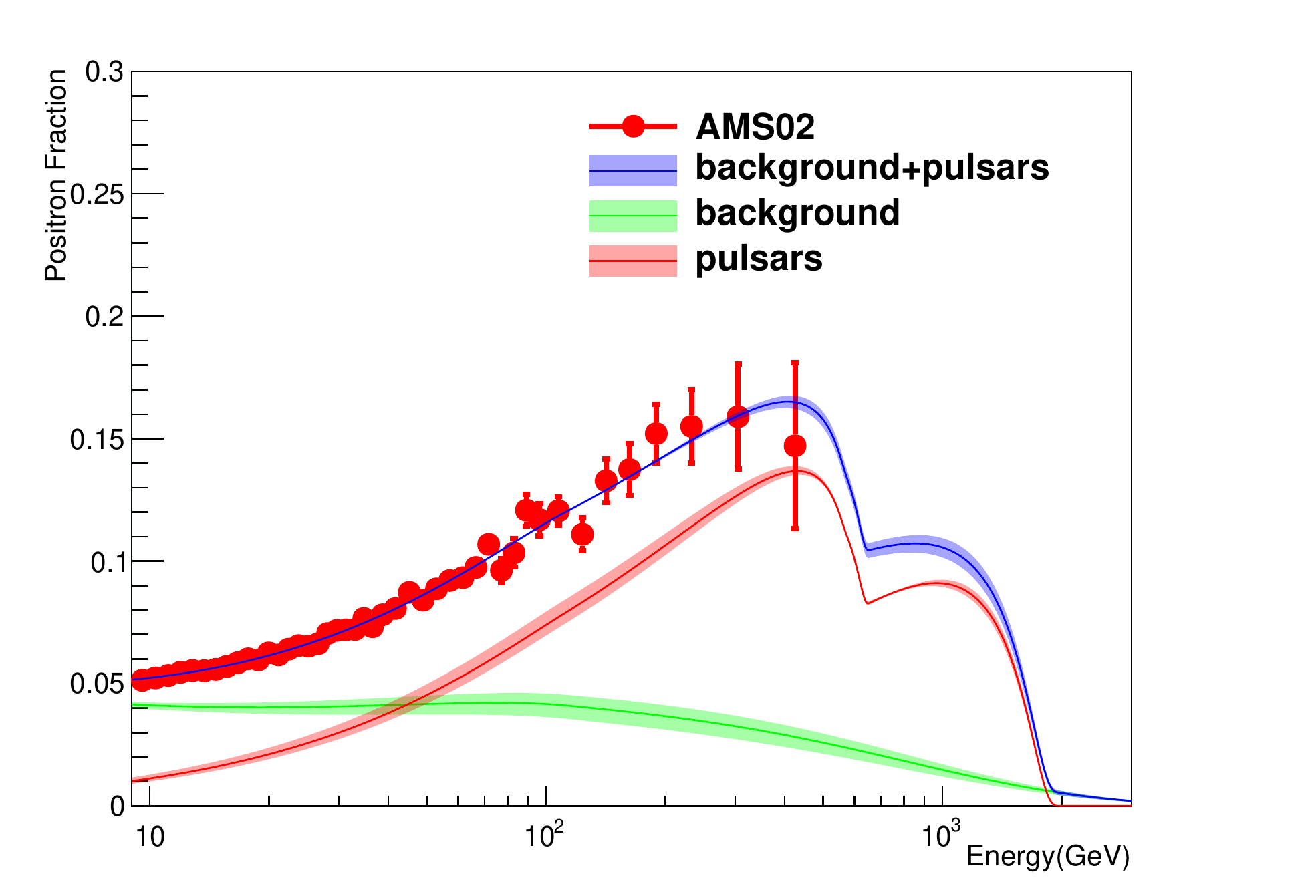}}
										   \subfloat[]{\includegraphics[width=0.5\textwidth]{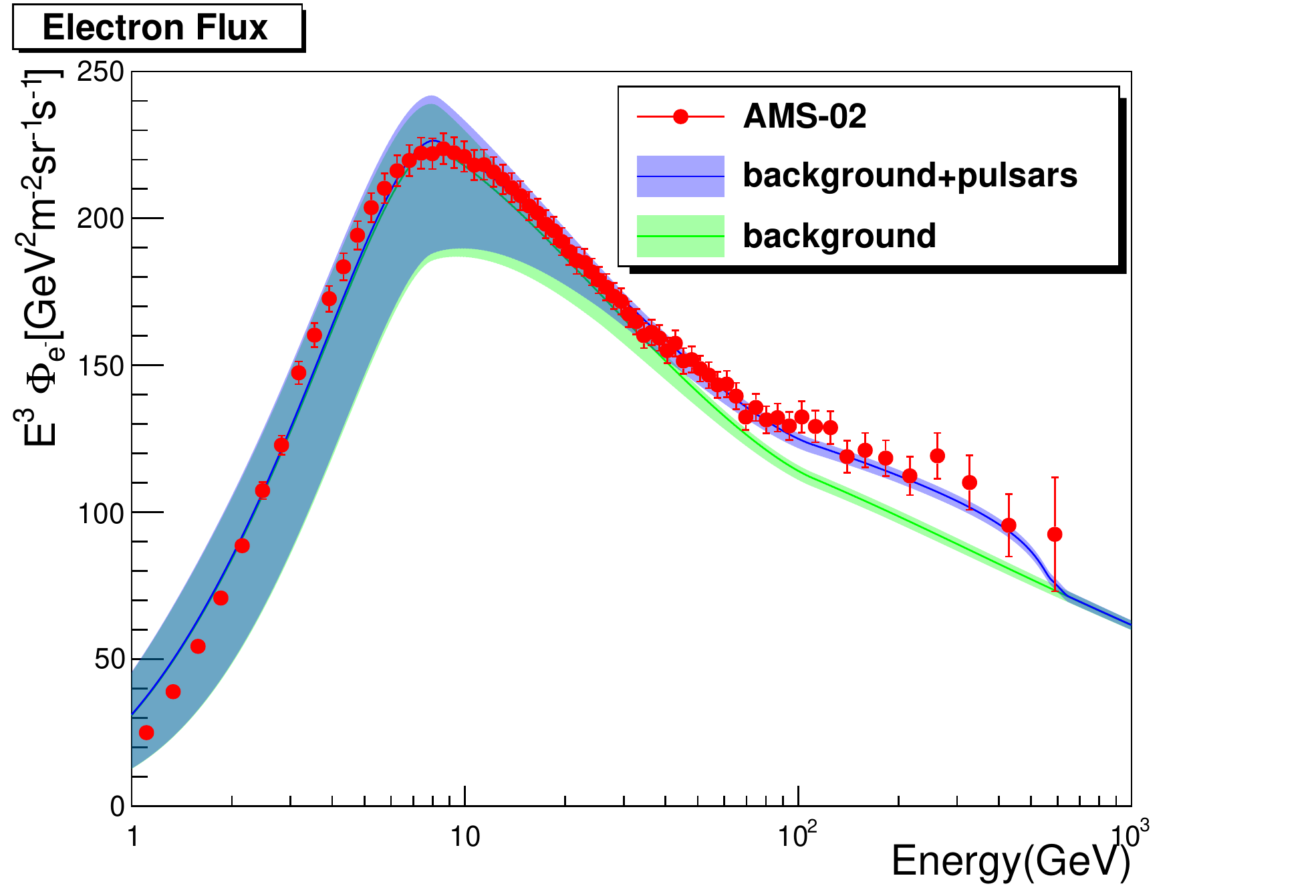}}\caption{Three pulsars fit to the positron fraction is presented here as an
											   example of multiple pulsars fit, as show in (a). Using the same parameters,
												   (b) shows the fitted parameters from positron fraction reproduce the
													   electron flux when the solar modulation potential 550 MV is applied
													   here. The error band in (a) shows the variation of propagation parameters within 95\% C.L., while that in (b) shows combined effect of propagation parameters and variation of solar modulation potential from 400 MeV to 800 MeV.}
													   \label{Fig:multi_electron_flux}%
													   \end{figure}

										   We choose three from the four ``surviving'' pulsars in the multiple
										   pulsars discussion. The input parameters are the age $T$, the distance
										   $d$ and the energy loss rate $\dot{E}$ of each pulsar while the
										   parameters we get from the fit is the normalization factor of primary
										   electron $A_{prim,e^{-}}$, the spectral index $\alpha$ and the
										   electron injection efficiency $\eta$. As shown in Fig.~\ref{Fig:multi_electron_flux}
										   (a), we obtain a good result from the multiple pulsar fit where $\chi{}^{2}/ndf=26.9/40$.
										   The parameters we get are $A{}_{prim,e^{-}}=0.50$, $\alpha=2.07$
										   and $\eta=2.58\%$.
The multiple pulsars interpretaion predicts a positron
fraction with a decrease up to 600 GeV and after that a bump up to
2000 GeV, which is possible to be observed with more accumulating AMS-02
data.

Using the parameters from the fit, we can reproduce the electron flux
measured by AMS-02 \cite{Aguilar:2014mma} in Fig.~\ref{Fig:multi_electron_flux}
(b). It shows that our electron background estimation in Section 2 $+$ pulsar contribution
matches the experimental data especially at high energies. Fig.~\ref{Fig:multi_electron_flux}
													   also shows that the effect of the uncertainty due to the propagation model is small at high energy. The solar
													   modulation potential is taken as 550 MV in the best fit result. The
													   solar modulation potential is varied between 400 MV and 800 MV to
													   show that its effect on low energy is quite large. To
													   reproduce the low energy electron flux more accurately, we need a
													   monthly low energy electron fluxes, which may be published by AMS collaboration to model solar modulation.

													   \section{Discussion and conclusion \label{sec-con}}

													   In this work, we investigate the possibility that the rise of the
													   positron fraction measured by AMS-02 can be explained by pulsars.
													   The propagation parameters and the injection spectrums of nuclei and
													   electrons are tuned according to the Boron-to-Carbon ratio and the
													   proton flux. It will be better to tune those parameters with Boron-to-Carbon
													   ratio and proton flux measured by AMS-02 since they are in the same
													   data taking period as the lepton fluxes. We find both the single
													   pulsar model and the multiple pulsar model can explain the AMS-02 data
													   very well. Six nearby pulsars are investigated as the single pulsar
													   sources of the high energy positrons and finally four survive from
													   all the conditions. The $\chi{}^{2}$s of these single pulsars in this work
													   are much smaller that those in \cite{Boudaud:2014dta}, mainly because we
													   set the cut-off energy equals to 5000 GeV while the authors of \cite{Boudaud:2014dta}
													   set it to 1000 GeV. With three mostly contributing pulsars, the multiple
													   pulsars model predicts a positron fraction with a decrease up to 600 GeV
													   and a bump up to 2000 GeV. For the low energy, a simple solar modulation
													   potential potential can not explain the measurement well. Thus, we
													   need the monthly electron fluxes which can describe solar activity
													   during the whole period.

													   It is shown that the positron excess measured by AMS-02 can be explained
													   by the pulsar scenario. Since the multiple pulsars can explain the experimental
													   data well, it will be difficult to exclude pulsar scenario by isotropy.
													   With accumulating AMS-02 data and future experiments, we can see the
													   positron fraction behavior up to higher energy which will either confirm or reject the
													   multiple pulsars scenario. If we consider other scenarios such as
													   Dark Matter, we have to look into other productions, antiproton for instance,
													   which have no contribution from pulsars.

													   \vspace{0.6cm}

													   \appendix
													   \section{Solving the diffusion energy-loss equation \label{app-A}}
To fix the notation and for the pedagogical purpose, in this appendix we give a brief review on solving the diffusion energy-loss equation ~\cite{Atoian:1995ux,Profumo:2008ms,Pochon2010,Kashiyama:2010ui,Yin:2013vaa}. 
													   The diffusion energy-loss equation is given by
													   \begin{eqnarray}
													   \left[\frac{\partial}{\partial t}-D(E)\nabla^2-\frac{\partial}{\partial E}b(E)\right]
f(\vec{x},t,E)
	=Q(\vec{x},t,E) \label{diffusion-energyloss-eq}
	\end{eqnarray}
	where $f(\vec{x},t,E)$ is the particle number density per unit energy interval, $D(E)>0$ is the diffusion coefficient, $b(E)\equiv -dE/dt>0$ is the energy loss rate, and $Q(\vec{x},t,E)$ is the source term.

	\subsection{Green function for the diffusion energy-loss equation}
	The Green function $G(\vec{x},t,E;\vec{x}_0,t_0,E_0)$ of \eqref{diffusion-energyloss-eq} is defined as
	\begin{eqnarray}
	\left[\frac{\partial}{\partial t}-D(E)\nabla^2-\frac{\partial}{\partial E}b(E)\right]
	G(\vec{x},t,E;\vec{x}_0,t_0,E_0)=\delta^3(\vec{x}-\vec{x}_0)\delta(t-t_0)\delta(E-E_0)
	\label{diffusion-energyloss-eq-Green-func}
	\end{eqnarray}
	The solution of \eqref{diffusion-energyloss-eq-Green-func} has been given in Ref.~\cite{Syrovatskii1959}. To fix the notation, let us briefly review the derivation.
	Define
	\begin{eqnarray}
	\phi(\vec{x},t,E;\vec{x}_0,t_0,E_0)\equiv b(E)G(\vec{x},t,E;\vec{x}_0,t_0,E_0)
	\end{eqnarray}
	Substituting $G=\phi/b$ into \eqref{diffusion-energyloss-eq-Green-func} gives
	\begin{equation}
	\left[\frac{\partial}{\partial t}-b(E)\frac{\partial}{\partial E}-D(E)\nabla^2\right]
	\phi(\vec{x},t,E;\vec{x}_0,t_0,E_0)
=b(E)\delta^3(\vec{x}-\vec{x}_0)\delta(t-t_0)\delta(E-E_0)
	\label{diffusion-energyloss-eq-Green-func-2}
	\end{equation}

	Let us make the variable transformation $(t,~E)\to(t',~\lambda)$ as follows,
	\begin{eqnarray}
	&& t'\equiv t-\tau(E,E_0)\;,\quad\text{with}\quad \tau(E,E_0)\equiv\int_E^{E_0}\frac{d\tilde{E}}{b(\tilde{E})}\label{t-prime}\\
		&& \lambda(E,E_0) \equiv \int_E^{E_0}d\tilde{E}\frac{D(\tilde{E})}{b(\tilde{E})}\label{lambda}
		\end{eqnarray}
		The Jacobian matrix of this transformation is easily obtained as
		\begin{eqnarray}
		\frac{\partial(t',~\lambda)}{\partial(t,~E)}\equiv
		\begin{pmatrix}
		\frac{\partial t'}{\partial t} & \frac{\partial t'}{\partial E}\\
			\frac{\partial \lambda}{\partial t} & \frac{\partial \lambda}{\partial E}
			\end{pmatrix}
			=\begin{pmatrix}
			1 & \frac{1}{b(E)}\\
				0 & -\frac{D(E)}{b(E)}
				\end{pmatrix} \label{Jacobian}
				\end{eqnarray}
				whose inverse is
				\begin{eqnarray}
				\frac{\partial(t,~E)}{\partial(t',~\lambda)}\equiv
				\begin{pmatrix}
				\frac{\partial t}{\partial t'} & \frac{\partial t}{\partial \lambda}\\
					\frac{\partial E}{\partial t'} & \frac{\partial E}{\partial \lambda}
					\end{pmatrix}
					=\begin{pmatrix}
					1 & \frac{1}{D(E)}\\
						0 & -\frac{b(E)}{D(E)}
						\end{pmatrix}
						\end{eqnarray}
						Thus,
						\begin{eqnarray}
						\frac{\partial}{\partial \lambda}=
						\frac{\partial t}{\partial \lambda}\frac{\partial}{\partial t}
						+\frac{\partial E}{\partial \lambda}\frac{\partial}{\partial E}
						=\frac{1}{D(E)}\left[\frac{\partial}{\partial t}-b(E)\frac{\partial}{\partial E}\right]
						\end{eqnarray}
						Substituting the relation $\partial_t-b(E)\partial_E=D(E)\partial_\lambda$ into \eqref{diffusion-energyloss-eq-Green-func-2} implies
						\begin{eqnarray}
						\left(\frac{\partial}{\partial \lambda}-\nabla^2\right)\phi
						&=&\frac{b(E)}{D(E)}\delta^3(\vec{x}-\vec{x}_0)\delta(t-t_0)\delta(E-E_0)\nonumber\\
						  &=&\frac{b(E_0)}{D(E_0)}\delta^3(\vec{x}-\vec{x}_0)\delta(t-t_0)\delta(E-E_0)
						  \label{diffusion-energyloss-eq-Green-func-3}
						  \end{eqnarray}

						  It follows from eq.~\eqref{Jacobian} that
						  \begin{eqnarray}
						  \det\left[\frac{\partial(t',~\lambda)}{\partial(t,~E)}\right]=-\frac{D(E)}{b(E)}
						  \end{eqnarray}
						  which implies
						  \begin{eqnarray}
\delta(t-t_0)\delta(E-E_0)
	&=&\left|\frac{D(E_0)}{b(E_0)}\right|\delta(t'-t_0)\delta(\lambda)\nonumber\\
	  &=&\frac{D(E_0)}{b(E_0)}\delta(t'-t_0)\delta(\lambda) \label{delta-function-relation}
	  \end{eqnarray}
	  where in the second equality we have used the properties: $D(E_0)>0$ and $b(E_0)>0$. Substituting \eqref{delta-function-relation} into \eqref{diffusion-energyloss-eq-Green-func-3}, we obtain
	  \begin{eqnarray}
	  \left(\frac{\partial}{\partial \lambda}-\nabla^2\right)\phi
	  =\delta(t'-t_0)\delta^3(\vec{x}-\vec{x}_0)\delta(\lambda)
	  \label{diffusion-energyloss-eq-Green-func-4}
	  \end{eqnarray}

	  As is well known, the Green function, which can be written as $G(\vec{x}-\vec{x}_0,\lambda)$ in
	  the spherically asymmetric approximation, of the diffusion equation satisfies
	  \begin{eqnarray}
\left(\frac{\partial}{\partial \lambda}-\nabla^2\right)G(\vec{x}-\vec{x}_0,\lambda)
=\delta^3(\vec{x}-\vec{x}_0)\delta(\lambda)
	\label{diffusion-eq}
	\end{eqnarray}
	with $\lambda\geq 0$. The solution of \eqref{diffusion-eq} is
	\begin{eqnarray}
G(\vec{x}-\vec{x}_0,\lambda)
	=\frac{1}{(4\pi\lambda)^{3/2}}\exp\left[-\frac{(\vec{x}-\vec{x}_0)^2}{4\lambda}\right]
	\end{eqnarray}

	Comparing \eqref{diffusion-energyloss-eq-Green-func-4} with \eqref{diffusion-eq}, we can read off the solution of $\phi$ as follows
	\begin{eqnarray}
	\phi&=&\delta(t'-t_0)
	\frac{1}{(4\pi\lambda)^{3/2}}\exp\left[-\frac{(\vec{x}-\vec{x}_0)^2}{4\lambda}
	\right]\nonumber\\
		&=&\delta[t-t_0-\tau(E,E_0)]
		\frac{1}{(4\pi\lambda)^{3/2}}\exp\left[-\frac{(\vec{x}-\vec{x}_0)^2}{4\lambda}
		\right]
		\end{eqnarray}
		where in the second equality we have used \eqref{t-prime}. Thus, we finally get the solution of \eqref{diffusion-energyloss-eq-Green-func} as follows
		\begin{eqnarray}
		G(\vec{x},t,E;\vec{x}_0,t_0,E_0)=\frac{\phi}{b(E)}=
		\frac{\delta[t-t_0-\tau(E,E_0)]}{b(E)[4\pi\lambda(E,E_0)]^{3/2}}
		\exp\left[-\frac{(\vec{x}-\vec{x}_0)^2}{4\lambda(E,E_0)}
		\right]\label{green-function}
		\end{eqnarray}
		where the functions $\tau(E,E_0)$ and $\lambda(E,E_0)$ are defined in eqs.~\eqref{t-prime} and \eqref{lambda}, respectively.

		\subsection{Solution for a burst-like source}

		Once we know the Green function, eq.~\eqref{green-function}, we can write down the solution of eq.~\eqref{diffusion-energyloss-eq} for a generic source as follows
		\begin{eqnarray}
		f(\vec{x},t,E)=\int d^3\vec{x'}dt'dE'G(\vec{x},t,E;\vec{x'},t',E')Q(\vec{x'},t',E')
		\label{solution-general}
		\end{eqnarray}
		In particular, for a burst-like source, the source function is proportional to $\delta^3(\vec{x}-\vec{x}_0)\delta(t-t_0)$, that is,
		\begin{eqnarray}
		Q(\vec{x},t,E)=Q(E)\delta^3(\vec{x}-\vec{x}_0)\delta(t-t_0) \label{source-burst-like}
		\end{eqnarray}
		where $Q(E)$ is an arbitrary function of $E$, $\vec{x}_0$ is the position of the source, and $t_0$ is the instantaneous time when the source bursts. Substituting eqs.~\eqref{green-function} and \eqref{source-burst-like} into eq.~\eqref{solution-general}, we obtain
		\begin{eqnarray}
		&&f(\vec{x},t,E)\nonumber\\
			&&=\int d^3\vec{x'}dt'dE'
			\frac{\delta[t-t'-\tau(E,E')]}{b(E)[4\pi\lambda(E,E')]^{3/2}}
			\exp\left[-\frac{(\vec{x}-\vec{x'})^2}{4\lambda(E,E')}
			\right]
			Q(E')\delta^3(\vec{x'}-\vec{x}_0)\delta(t'-t_0)\nonumber\\
				&&=\int dE'\frac{\delta[t-t_0-\tau(E,E')]}{b(E)[4\pi\lambda(E,E')]^{3/2}}
				\exp\left[-\frac{(\vec{x}-\vec{x}_0)^2}{4\lambda(E,E')}
				\right]
				Q(E')\label{solution-general-2}
				\end{eqnarray}

				Denote the solution of the equation
				\begin{eqnarray}
				0=t-t_0-\tau(E,E')=t-t_0-\int_E^{E'}\frac{d\tilde{E}}{b(\tilde{E})}\label{E0-equation}
				\end{eqnarray}
				is $E'=E_0$, then we have
				\begin{eqnarray}
				\delta[t-t_0-\tau(E,E')]=b(E_0)\delta(E'-E_0) \label{delta-function-relation-2}
				\end{eqnarray}
				where we have used
				\begin{eqnarray}
				\left|\frac{d\tau}{dE'}\right|_{E'=E_0}\delta[t-t_0-\tau(E,E')]=\delta(E'-E_0)
				\end{eqnarray}

				Substituting eq.~\eqref{delta-function-relation-2} into eq.~\eqref{solution-general-2}, we have
				\begin{eqnarray}
				f(\vec{x},t,E)&=&\int dE'\frac{b(E_0)\delta(E'-E_0)}{b(E)[4\pi\lambda(E,E')]^{3/2}}
				\exp\left[-\frac{(\vec{x}-\vec{x}_0)^2}{4\lambda(E,E')}
				\right]Q(E')\nonumber\\
					&=&\frac{Q(E_0)}{[4\pi\lambda(E,E_0)]^{3/2}}\frac{b(E_0)}{b(E)}
					\exp\left[-\frac{(\vec{x}-\vec{x}_0)^2}{4\lambda(E,E_0)}
					\right] \label{solution-general-3}
					\end{eqnarray}
					where the initial energy $E_0$ is defined as the solution of eq.~\eqref{E0-equation}. In other words, if we know $t-t_0$ and $E$, we can find $E_0$ by solving the equation
					\begin{eqnarray}
					t-t_0=\tau(E,E_0)=\int_E^{E_0}\frac{d\tilde{E}}{b(\tilde{E})} \label{E0-equation-2}
					\end{eqnarray}

					Define the diffusion distance $r_{\mathrm{dif}}$ as
					\begin{eqnarray}
					r_{\mathrm{dif}}(E,E_0)\equiv\sqrt{4\lambda(E,E_0)}
					=2\sqrt{\int_E^{E_0}d\tilde{E}\frac{D(\tilde{E})}{b(\tilde{E})}}
					\end{eqnarray}
					we can rewrite eq.~\eqref{solution-general-3} as
					\begin{eqnarray}
					f(\vec{x},t,E)=\frac{Q(E_0)}{\pi^{\frac{3}{2}}r_{\mathrm{dif}}^3}
\frac{b(E_0)}{b(E)}\exp\left(-\frac{r^2}{r_{\mathrm{dif}}^2}\right)
	\label{solution-general-4}
	\end{eqnarray}
	with $r^2\equiv (\vec{x}-\vec{x}_0)^2$.

	\subsection{Solution for a burst-like source with power-law spectrum}

	Consider the case when the function $Q(E)$ in eq.~\eqref{source-burst-like} is a power-law function, that is,
	\begin{eqnarray}
Q(E)=Q_0E^{-\alpha}\exp\left(-\frac{E}{E_{\mathrm{cut}}}\right)
	\end{eqnarray}
	where $Q_0$ is a normalization constant, $\alpha$ is the spectral index, and $E_{\mathrm{cut}}$ is the cutoff energy. The diffusion coefficient $D(E)$ and the energy loss rate $b(E)$ are assumed to take the form
	\begin{eqnarray}
	&&D(E)=\beta D_0\left(\frac{E}{E_0}\right)^\delta \label{D-E}\\
	       &&b(E)=b_0E^2 \label{b-E}
	       \end{eqnarray}
	       where $\beta\equiv v/c$ is the ratio of velocity to speed of light, the constant $D_0$ and the index $\delta$ can be figured out by the background fitting in Sec.~\ref{sec-background}, the constant $b_0$ is given in Sec.~\ref{sec-single}, and $E_0$ should be determined by eq.~\eqref{E0-equation-2}.

	       Let us calculate the initial energy $E_0$ first. Denote the diffusion time $t_{\mathrm{dif}}\equiv t-t_0$. It follows from eqs.~\eqref{E0-equation-2} and \eqref{b-E} that
	       \begin{eqnarray}
	       t_{\mathrm{dif}}=\int_E^{E_0}\frac{d\tilde{E}}{b_0\tilde{E}^2}
=\frac{1}{b_0}\left(\frac{1}{E}-\frac{1}{E_0}\right)
	\end{eqnarray}
	which implies
	\begin{eqnarray}
	\frac{1}{E}-\frac{1}{E_0}=b_0t_{\mathrm{dif}}
	\end{eqnarray}
	from which, we see $\frac{1}{E}>b_0t_{\mathrm{dif}}$, that is, $E<\frac{1}{b_0t_{\mathrm{dif}}}$. Denote $E_{\mathrm{max}}\equiv 1/(b_0t_{\mathrm{dif}})$, then
	\begin{eqnarray}
	\frac{1}{E}-\frac{1}{E_0}=\frac{1}{E_{\mathrm{max}}}
	\end{eqnarray}
	which gives
	\begin{eqnarray}
	E_0=\frac{EE_{\mathrm{max}}}{E_{\mathrm{max}}-E}
	\end{eqnarray}
	Thus, we obtain the following pieces in eq.~\eqref{solution-general-4}:
	\begin{eqnarray}
Q(E_0)&=&Q_0E_0^{-\alpha}\exp\left(-\frac{E_0}{E_{\mathrm{cut}}}\right)
	=Q_0\left(\frac{EE_{\mathrm{max}}}{E_{\mathrm{max}}-E}\right)^{-\alpha}
	\exp\left(-\frac{EE_{\mathrm{max}}}{E_{\mathrm{cut}}(E_{\mathrm{max}}-E)}\right)\nonumber\\
		&=&Q_0E^{-\alpha}\left(1-\frac{E}{E_{\mathrm{max}}}\right)^\alpha
		\exp\left[-\frac{E/E_{\mathrm{cut}}}{(1-E/E_{\mathrm{max}})}\right]\label{Q-E0}
		\end{eqnarray}
		\begin{eqnarray}
		\frac{b(E_0)}{b(E)}=\frac{E_0^2}{E^2}=\left(1-\frac{E}{E_{\mathrm{max}}}\right)^{-2}
		\label{b-E0-to-b-E}
		\end{eqnarray}
		Substituting eqs.~\eqref{Q-E0} and \eqref{b-E0-to-b-E} into eq.~\eqref{solution-general-4}, we obtain
		\begin{eqnarray}
		f(\vec{x},t,E)=\frac{Q_0E^{-\alpha}}{\pi^{\frac{3}{2}}r_{\mathrm{dif}}^3}
		\left(1-\frac{E}{E_{\mathrm{max}}}\right)^{\alpha-2}
		\exp\left[-\frac{E/E_{\mathrm{cut}}}{(1-E/E_{\mathrm{max}})}
		-\frac{r^2}{r_{\mathrm{dif}}^2}\right] \label{solution-final}
		\end{eqnarray}
		which is consistent with previous works (for example, eq.~(6) of Ref.~\cite{Yin:2013vaa}).

		Now let us figure out the diffusion distance $r_{\mathrm{dif}}$. To this end, we substitute eqs.~\eqref{D-E} and \eqref{b-E} into eq.~\eqref{lambda}, and get
		\begin{eqnarray}
		\lambda(E,E_0)&=&\int_E^{E_0}d\tilde{E}\frac{\beta D_0\left(\frac{\tilde{E}}{E_0}\right)^\delta}{b_0\tilde{E}^2}
		=\frac{\beta D_0}{E_0^\delta b_0}\int_E^{E_0}d\tilde{E}\tilde{E}^{\delta-2}\nonumber\\
		 &=&\frac{\beta D_0}{E_0^\delta b_0(\delta-1)}
\left(E_0^{\delta-1}-E^{\delta-1}\right)
	\end{eqnarray}
	which can also be written as
	\begin{eqnarray}
	\lambda(E,E_0)=\beta D_0\left(\frac{E}{E_0}\right)^\delta
	\frac{t_{\mathrm{dif}}}{b_0t_{\mathrm{dif}}(\delta-1)}
	\frac{1}{E}\left[\left(\frac{E_0}{E}\right)^{\delta-1}-1\right]
	\end{eqnarray}
	Comparing the above equation with eq.~\eqref{D-E}, $E_{\mathrm{max}}\equiv 1/(b_0t_{\mathrm{dif}})$ and $E_0/E=(1-E/E_{\mathrm{max}})^{-1}$, we obtain
	\begin{eqnarray}
	\lambda(E,E_0)=\frac{D(E)t_{\mathrm{dif}}}{(1-\delta)}
	\frac{E_{\mathrm{max}}}{E}
	\left[1-\left(1-\frac{E}{E_{\mathrm{max}}}\right)^{1-\delta}\right]
	\end{eqnarray}
	Thus, the diffusion distance $r_{\mathrm{dif}}$ is given by
	\begin{eqnarray}
	r_{\mathrm{dif}}\equiv 2\sqrt{\lambda(E,E_0)}
	=2\sqrt{\frac{D(E)t_{\mathrm{dif}}}{(1-\delta)}
		\frac{E_{\mathrm{max}}}{E}
		\left[1-\left(1-\frac{E}{E_{\mathrm{max}}}\right)^{1-\delta}\right]}
		\label{diff-distance-final}
		\end{eqnarray}
		which is consistent with previous works (for example, eq.~(10) of Ref.~\cite{Hooper:2008kg} and eq.~(7) of Ref.~\cite{Yin:2013vaa}).

		\section*{Acknowledgments}

		We would like to thank M.~Boudaud, S.~Caroff, S.~J.~Lin, A.~Putze, S.~Rosier-Lees and P.~Salati for helpful discussions.
		This work is supported in part by the National Natural Science Foundation
		of China (NSFC) under Grant Nos. 11375277, 11410301005 and 11005163,  the Fundamental
		Research Funds for the Central Universities, and Sun Yat-Sen
		University Science Foundation.

		\bibliographystyle{unsrt}
		\bibliography{pulsar}

		\end{document}